\documentclass[11pt]{article}
\usepackage[a4paper, margin= 1in]{geometry}
\usepackage{amsmath,amssymb}
\usepackage{lipsum}
\usepackage{subfig}
\usepackage{graphicx}% Include figure files
\usepackage{slashed}
\usepackage{dcolumn}% Align table columns on decimal point
\usepackage{bm}% bold math
\usepackage{hyperref}% add hypertext capabilities
\usepackage{physics}
\usepackage{authblk}
\usepackage{xcolor}
\usepackage[maxfloats=256]{morefloats}
\usepackage{ragged2e}
\usepackage{comment} 
\usepackage{dcolumn}% Align table columns on decimal point
\usepackage{bm}%
\usepackage{float}	
\usepackage[T1]{fontenc} % if needed
\usepackage{color}
\usepackage{mdframed}

\usepackage{textcomp}
\usepackage{mathtools, cuted}
\usepackage{enumerate}
\usepackage{mathrsfs}
\usepackage{verbatim}
\usepackage{slashed}
\usepackage{epsfig}
\usepackage{enumerate}
\usepackage{wrapfig}
\usepackage{amsfonts}
\usepackage{multirow}
\usepackage{tikz}
\usepackage{amsmath}
\usepackage[normalem]{ulem}
\usepackage[normalem]{ulem}

\newcommand{\bea}{\begin{eqnarray}}
	\newcommand{\eea}{\end{eqnarray}}
\newcommand{\be}{\begin{equation}}
	\newcommand{\ee}{\end{equation}}
\newcommand{\bal}{\begin{align}}
	\newcommand{\ealign}{\end{align}}
\newcommand{\eal}{\end{align}}
\newcommand{\ben}{\begin{enumerate}}
\newcommand{\een}{\end{enumerate}}
\newcommand{\bit}{\begin{itemize}}

\newcommand{\nn}{\nonumber}

\newcommand{\bd}{\begin{displaymath}}
\newcommand{\ed}{\end{displaymath}}

\newcommand{\mhpm}{\ensuremath{m_{H^{\pm}} \,\, }}

\newcommand{\vsm}{\ensuremath{v_{_{\! \rm SM}}} }
\newcommand{\vs}{\ensuremath{v_{_{\! \rm s}}} }

\title{\textbf{Higgs Decay Signal Strengths in an Extended 2HDM}}

\author[1]{Hrishabh Bharadwaj\thanks{\textit{hrishabhphysics@gmail.com}}}
\author[2]{Mamta Dahiya\thanks{Corresponding author: \textit{mamta.dahiya@sgtbkhalsa.du.ac.in}}}
\author[2,4]{Sukanta Dutta\thanks{\textit{sukanta.dutta@sgtbkhalsa.du.ac.in}} }
\author[3]{Ashok Goyal\thanks{\textit{agoyal45@yahoo.com } }}

\affil[1]{\textit{\small Government Mahila Degree College, Budaun, Mahatmaa Jyotiba Phule Rohilkhand University, Bareilly, 243601, Uttar Pradesh, India. }}
\affil[2]{\textit{\small SGTB Khalsa College, University of Delhi, Delhi, 110007, India}}
\affil[3]{\textit{\small Department of Physics and Astrophysics, University of Delhi, Delhi, 110007, India}}
\affil[4]{\textit{\small Delhi School of Analytics, Institution of Eminence, University of Delhi, Delhi, 110007, India}}

\date{}
\begin{document}
	\maketitle
\begin{abstract}
The combined CMS and ATLAS analysis of \(h \to Z\gamma\) reveals a \(2\sigma\) excess over the  Standard Model (SM) prediction, hinting at potential new physics. In contrast, the observed agreement of \(\mu_{\gamma\gamma}\) with the SM imposes stringent constraints on such extension. In this context, we investigate a Two Higgs Doublet Model (2HDM) extended by a complex scalar singlet and a vector-like lepton (VLL). This framework successfully accommodates the Higgs decay signal strengths \(\mu_{{}_{W \,W^\star}}\), \( \mu_{Z\gamma}\,\), and \(\, \mu_{\gamma\gamma}\) while also satisfying precision electroweak constraints,  the observed experimental value of muon \(g-2\),  and bounds from LEP II data. 
We identify a viable region of parameter space that is consistent with all low energy data and current experimental limits on the Higgs decay signal strengths. 
\end{abstract}
\vspace{1em}

%\end{frontmatter}
\section{Introduction}
\label{sec:intro}
The ATLAS and CMS collaborations recently reported the first evidence for $h \to Z\gamma$ decay based on their combined analysis of LHC Run 2 datasets, with integrated luminosities of 139 fb$^{-1}$  (ATLAS) and 138 fb$^{-1}$ (CMS) at $\sqrt{s}$ = 13 TeV. The improved combined measured signal strength is:
\bea
\mu_{Z\gamma}^{\text{combined}} &=& 2.2 \pm 0.7 \quad \text{\cite{ATLAS:2023yqk}}
\label{eq:muZG-combined}
\eea
showing a $1.9\sigma$ excess over the Standard Model (SM) expectation. However,  the signal strength for $h \rightarrow \gamma\gamma$ decay, 
\be
\mu_{\gamma\gamma} = 1.10 \pm 0.06 \quad \cite{ParticleDataGroup:2024cfk} 
\label{eq:muGG}
\ee
aligns well with the SM prediction, imposing strong constraints on new physics explanations for the   $\mu_{Z\gamma}$ excess. 

Since SM corrections at the two-loop level, including QCD and electroweak effects~\cite{Spira:1991tj,Belusca-Maito:2015lna,Gehrmann:2015dua,Buccioni:2023qnt,Chen:2024vyn, Sang:2024vqk}, are insufficient to account for this deviation, various BSM scenarios have been explored. These include multi-Higgs doublet models \cite{Panghal:2023iqd,Arhrib:2024wjj,Arhrib:2024itt,Chen:2024oru}, supersymmetry \cite{Hu:2024slu,Israr:2024ubp}, left-right symmetric models, ultra-light axion-like particles \cite{Cheung:2024kml}, effective field theory (EFT) approaches \cite{He:2024bxi,Mantzaropoulos:2024vpe}, and models with off-diagonal Higgs-$Z$-bosons couplings \cite{Boto:2023bpg}.

Several low-energy observables show deviations from SM predictions, notably the muon anomalous magnetic moment and the CDF measurement of the $W$-boson mass. The latest muon \(g-2\) measurement reports a  $5\sigma$ discrepancy:
\be
 \Delta a_\mu = a_\mu^{\text{exp}} - a_\mu^{\text{SM}} = (2.49 \pm 0.08) \times 10^{-9} \,\,\,\, \cite{Muong-2:2023cdq,AOYAMA20201}.
 \label{eq:MDMexpt}
 \ee
While the recent CMS measurement of $m_{W} $ \cite{CMS:2024nau} aligns with the SM, the CDF result \cite{CDF:2022hxs,ParticleDataGroup:2024cfk,LHC-TeVMWWorkingGroup:2023zkn,deBlas:2022hdk} shows a $7\sigma$ deviation, leaving the discrepancy unresolved.

To address the aforementioned discrepancies, we seek a unified framework capable of resolving these deviations. We build upon the scalar sector of the 2HDM as explored in \cite{Bharadwaj:2021tgp, Bharadwaj:2024gfo}, where a common parameter space was identified, incorporating constraints from the muon anomalous magnetic moment, Higgs boson decays, and LEP II results. In this work, we compute the signal strength of the $h \to Z\gamma$ decay and investigate whether the parameter space required to resolve the muon $(g-2)$ anomaly remains consistent with the observed $h \to Z\gamma$ signal strength. This analysis is performed within a 2HDM framework extended by a SM-singlet complex scalar and a  charged VLL. While previous studies incorporated VLL interactions, our model differs by omitting the Yukawa interactions between the VLL and right-handed SM leptons, making it distinct from the setup in \cite{Bharadwaj:2021tgp}.
 
This paper is organized as follows: In Section~\ref{sec:model}, we briefly describe the model. The computation of additional contributions to the Higgs decay channels $h \to \gamma\gamma$ and $h \to Z\gamma$ at the one-loop level in our model is discussed in Section~\ref{sec:signalstregth}. The numerical results for the parameter space region that simultaneously accommodates the measured signal strengths and the muon $(g-2)$ discrepancy are presented in Section~\ref{sec:num}. Finally, we summarize our findings in Section~\ref{sec:sum}.

\section{The Model}
\label{sec:model}
The extended 2HDM scalar sector  includes two SU(2)${}_L$ scalar doublets and a neutral singlet scalar $\Phi_3$. The doublet $\Phi_1$ and singlet $\Phi_3$  acquire real vacuum expectation values (VEVs) $\vsm$ and $\vs$, respectively. The scalar doublets $\Phi_1$, $\Phi_2$ and complex singlet $\phi_3$ are defined as
	\bea
	\Phi_1\equiv\left[\begin{array}{c} \phi_1^+\\
		\frac{1}{\sqrt{2}} \,\left(\vsm+\phi_1^0 + i\,\eta_1^0\right)\\\end{array}\right],\quad \Phi_2\equiv\left[\begin{array}{c} \phi_2^+\\
		\frac{1}{\sqrt{2}} \,\left( \phi_2^0 + i\,\eta_2^0\right)\\ \end{array}\right]
	\quad {\rm and} \quad  \,\Phi_3\equiv \frac{1}{\sqrt{2}}\, \left[\vs +\phi_3^0 + i\,\eta_3^0 \right],\nn\\ \label{scalarfields}
\eea
The scalar Lagrangian, ${\mathscr L}_{\rm scalar}$, consists of the kinetic terms for the scalar fields along with the scalar potential.
\bea
{\mathscr L}_{\rm scalar}&=& \sum_{i=1,2}\left(D_\mu\Phi_i\right)^\dagger\ \left(D^\mu\Phi_i\right)\  +\ \left(D_\mu\Phi_3\right)^\star\ \left(D_\mu\Phi_3\right) - \ V_{\rm scalar}, \quad \text{where}  \nn\\
	V_{\rm scalar}
&=& -\frac{1}{2} m_{11}^2  \left(\Phi_1^\dagger\Phi_1\right)  - \frac{1}{2} m_{22}^2  \left(\Phi_2^\dagger\Phi_2\right) 
 -\frac{1}{2} m_{33}^2\ \Phi_3^*\Phi_3
+ \frac{\lambda_1}{2} \left(\Phi_1^\dagger\Phi_1\right)^2 +\frac{\lambda_2}{2} \left(\Phi_2^\dagger\Phi_2\right)^2\nn\\
&&  + \lambda_3 \left(\Phi_1^\dagger\Phi_1\right)  \left(\Phi_2^\dagger\Phi_2\right) + \lambda_4 \left(\Phi_1^\dagger\Phi_2\right) \left(\Phi_2^\dagger\Phi_1\right) + \frac{1}{2} \left[\lambda_5\  \left(\Phi_1^\dagger\Phi_2\right)^2 +\ h.c. \right]    \nn\\
&& 
+ \frac{\lambda_8}{2} \left( \Phi_3^*\Phi_3 \right)^2+ \lambda_{11} \left\vert \Phi_1\right\vert^2 \Phi_3^*\Phi_3 
+ \lambda_{13} \left\vert\Phi_2\right\vert^2 \Phi_3^*\Phi_3
  -i\, \kappa\,\, \left[\left( \Phi_1^\dagger\Phi_2 + \Phi_2^\dagger \Phi_1 \right)\,\,\left(\Phi_3-\Phi_3^\star\right)  \right] \nn\\ 
\label{eq:scalarpot}
\eea
A discrete $Z_2$ symmetry is imposed to forbid direct interactions of $\Phi_2$ doublet with  the SM fields and the $\Phi_1$ doublet. Under this symmetry, $\Phi_2$ and  $\Phi_3$ are assigned odd parity, whereas the SM  fields and $\Phi_1$ are even. Additionally, a global $U(1)$ symmetry is introduced under which $\Phi_3 \to e^{i\alpha} \Phi_3$, in order to reduce the number of independent parameters in the scalar potential. This symmetry is allowed to be softly broken by the $\kappa$ term.

The mass matrices for neutral scalars and pseudoscalars remain decoupled due to the absence of mixing between the imaginary component of the doublet $\Phi_2$ and the real component of either $\Phi_1$ or $\Phi_3$. The CP-even neutral scalar mass matrix is a $2 \times 2$ matrix that arises from the mixing of the real components of the SM-like doublet $\Phi_1$ and the singlet $\Phi_3$. Diagonalizing this CP-even mass matrix using an orthogonal rotation matrix parameterized by the mixing angle $\theta_{13}$ yields two mass eigenstates, $h_1$ and $h_3$. Similarly, diagonalizing the CP-odd scalar mass matrix for $\eta_2^0$ and $\eta_3^0$ using an orthogonal rotation matrix parameterized by the mixing angle $\theta_{23}$ results in the pseudoscalar mass eigenstates $A^0$ and $P^0$. For explicit expressions of the mass eigenstates and the relationships between mass parameters and scalar couplings in the Lagrangian, the reader is referred to \cite{Bharadwaj:2021tgp, Bharadwaj:2024gfo}.

Among the remaining neutral and charged scalar mass eigenstates, $\eta_1^0$ and $\phi_1^\pm$ correspond to the massless Nambu-Goldstone bosons, while the scalars $\phi_2^0$ and $\phi_2^\pm$ are now relabeled as $h_2$ and $H^\pm$, respectively.

The Yukawa terms for SM fields are given by:
\bea
	-{\mathscr L}^\prime_{\rm Yukawa} &=& y_{u}\ \overline{Q_L}\ \widetilde{\Phi_1}\ u_R\ +\ y_{d}\ \overline{Q_L}\ \Phi_1\ d_R\ +\ y_{_\ell}\ \overline{\ell_L}\ \Phi_1 \ e_R\  + y_1\ \overline{\ell_L}\ \Phi_2 \ e_R  +\ \text{h.c.}. \label{SMYukawa}
	\eea
where \(Q_L\) and \(\ell_L\) are the quark and lepton doublets under \(SU(2)_L\) gauge group. Here, the $Z_2$ symmetry is allowed to get softly broken in the Yukawa interactions involving $\Phi_2$ to allow the interactions between the  SM leptons with CP-odd pseudoscalars. 

In addition to the scalar sector, the model includes a  singlet chiral vector-like leptons, where the left-handed (right-handed) VLL is odd (even) under $Z_2$ symmetry. The Lagrangian for VLL  is given by
	\bea
	{\mathscr L}_{\rm VLL}&=&  \overline{\chi}\ i \left(\not \! {\partial} - i g^\prime\frac{Y}{2} \not \! \! B \right) \chi\ -\ m_{\chi}\ \overline{\chi}\ \chi -  y_2\ \overline{\chi_L}\ \chi_R\ \Phi_3 
	 \label{VLYukawa}
	\eea
In contrast to \cite{Bharadwaj:2021tgp}, our model excludes the $y_3$-induced Yukawa interactions, which previously facilitated mixing between SM fermions and the VLL. In this framework, the global $U(1)$ symmetry is allowed to be softly broken by the Yukawa coupling $y_2$.

The Yukawa interactions given in \eqref{SMYukawa} and \eqref{VLYukawa}  can be re-written as:
\begin{eqnarray}
-{\mathscr L}^{\rm \small Yukawa}_{\rm \small SM\, Fermions}\!\!\!\!&=&\!\!\!\!\sum_{s_i\equiv h_1,h_3}\frac{y_{ffs_i}}{\sqrt{2}} \left(v_{\rm SM}\,\,\delta_{s_i,h_1} + s_i\right) \bar f\ f + \frac{y_{llh_2}}{\sqrt{2}} h_2 \bar l^- l^-  \nn \\ 
&&  + \sum_{s_i\equiv P^0,A^0} \frac{y_{lls_i}}{\sqrt{2}} s_i \bar l^-  \gamma_5\ l^- + \left[y_{l\nu H^-}\  \bar \nu_l\ P_R\ l^-   H^+ +  {\rm h.c.} \right],
%\nn\\
%&& 
 \label{eq:yukawa-sm}\\
 		-{\mathscr L}^{\rm Yukawa}_{\rm VL\ Leptons}&=&
 		\sum_{s_i\equiv h_1,h_3,A^0,P^0}\frac{1}{\sqrt{2}} \left(v_s\,\, \delta_{s_i,h_3}+ s_i\right)\ \bar \chi \left(y_{\chi\chi s_i} P_R+  y_{\chi\chi s_i}^\star P_L\right) \chi,
 			\label{eq:yukawa-VLL}
 	\end{eqnarray}
where $f$ and $l^-$ denote SM fermions and SM charged leptons, respectively, and 
$y^\prime$s represent the couplings of the fermionic fields ($f, l, \nu, \chi$) with the charged or neutral scalar/ pseudoscalar mass eigenstates. These Yukawa couplings depend on the vacuum expectation values (VEVs), the parameters $y_i$ (for $i=1,2$), and the mixing angles $\theta_{13}$ and $\theta_{23}$. The explicit expressions for these couplings can be found in Table 1 of \cite{Bharadwaj:2021tgp}.

There are 12 parameters in the scalar potential~\eqref{eq:scalarpot}, namely,
$\vs, m_{11}$, $ m_{22}$, $m_{33}$, $\lambda_{i= 1, 3, 4, 5, 8, 11,13}$  and $ \kappa$. Among these, the parameter $\lambda_2$ does not contribute to the mass spectrum and is therefore not constrained by our analysis. Additionally, $\vs$ and $m_{22}$ are not directly constrained. The eight dimensionless couplings in the scalar potential and Yukawa sector are assumed to be real to preserve CP invariance. Thus, along with the Yukawa couplings $y_i$, we consider the effects of the following 13 physical parameters in our study:
 \bea
 {\text{Masses}}&:& m_{h_1},\, m_{h_2},\, m_{h_3},\, m_{H^\pm},\,\ m_{A^0},\, m_{P^0}, \,\, m_\chi \nn\\
 {\text{Mixing Angles}}&:& \theta_{13},\, \theta_{23} \nn\\
 {\text{Couplings}}&:&  y_1, \, y_2, \, \lambda_{{h_1 H^+ H^-}},\, \lambda_{{h_3 H^+ H^-}}
 \label{eq:parameters2}
 \eea
 where \(\lambda_{h_1 H^+ H^-}\) and \(\lambda_{h_3 H^+ H^-}\) are the dimensionless scalar triple couplings of the charged Higgs bosons with neutral scalars, given by 
	\bea 
	\lambda_{h_1 H^+ H^-}  = \lambda_3 \,\cos\theta_{13} +  \frac{v_s}{\vsm} \,\lambda_{13}\,\sin\theta_{13}
	% \label{eq:gh1hphm}\\
	\quad \text{and} \quad
	\lambda_{h_3 H^+ H^-}  =   \frac{v_s}{\vsm} \lambda_{13}  \cos\theta_{13}\ -\ \lambda_3\  \sin\theta_{13}	 
	\label{eq:ghihphm}
	\eea
The co-positivity conditions  impose mutually exclusive allowed regions for $\lambda_4$ and $\lambda_5$ \cite{Bharadwaj:2021tgp, Bharadwaj:2024gfo}, leading to :
\bea
  \Theta(\left\vert \lambda_5\right\vert-\lambda_4)=\left\{\begin{array}{cc} \Theta\left[m_{H^\pm}^2-(m_{A^0}^2+m_{P^0}^2)\right] &\,\,\text{for}\,\,\,\, m_{h_2}^2>m_{A^0}^2+m_{P^0}^2 \\
   \!\!\!\!\!\!\Theta\left[m_{h_2}^2-m_{H^\pm}^2\right]\,\,\,\,\,\,\,\,\,\,\,\,\,\,\,
     & \,\,\text{for}\,\,\,\, m_{h_2}^2<m_{A^0}^2+m_{P^0}^2 \\
  \end{array}\right. 
 \label{derivedpositivity}
  \eea
In this article, we explore the  phenomenologically interesting  region where $m_{H^\pm}^2$ and $m_{h_2}^2$  are  $>m_{A^0}^2+m_{P^0}^2$. For further analysis,  all Yukawa couplings are restricted to the range \( \left\vert y_i\right\vert \le \!\sqrt{4\,\pi}\) as required by tree-level perturbative unitarity. 
\section{Higgs Decay Signal Strengths}
\label{sec:signalstregth}
Any multi-Higgs model must accommodate an SM-like Higgs boson with the mass and signal strengths measured at the LHC. The LHC data favor a scalar eigenstate $h^{\rm SM}$ with a mass of approximately $125$ GeV~\cite{ParticleDataGroup:2024cfk}. This motivates us to identify the lightest CP-even neutral scalar, $h_1$, with the observed Higgs boson and set $m_{h_1} = 125$ GeV. Furthermore, the couplings of $h_1$ to fermions and gauge bosons are the corresponding SM Higgs couplings, suppressed by a factor of $\cos\theta_{13}$ due to $\Phi_1-\Phi_3$ mixing. 

Signal strength is a key observable used to quantify the deviation of the observed signal from the SM expectation in specific decay channels. To constrain the parameter space of the model through Higgs decays, we require that $h_1$ decays account for the measured total Higgs decay width. For this purpose, we define the signal strength \(\mu_{_{XY}}\) relative to $h^{\rm SM}$ production via the dominant gluon fusion process in  \(p-p\) collisions, followed by its decay into \(X, Y\) pairs, under the narrow width approximation:
\begin{equation} \mu_{{}{XY}} = \frac{\sigma(pp\to h_1\to XY)}{\sigma(pp\to h\to XY)^{\textrm{SM}}} = \frac{\Gamma\left( h_1\to g g\right)}{\Gamma \left(h^{SM} \to g g\right)} \times \frac{\textrm{BR}\left(h_1\to X Y\right)}{\textrm{BR}\left(h^{\rm SM}\to X Y\right)}
= \cos^2\theta_{13} \times \frac{\textrm{BR}\left(h_1\to X Y\right)}{\textrm{BR}\left(h^{\rm SM}\to X Y\right)}.
\label{eq:signalstrength} \end{equation}
\subsection{$h_1 \to WW^\star$}
\label{subsec:mugg}
The partial decay width of $h_1\to W\,W^\star$ channel in our model is related to the corresponding SM value  as
\bea
\Gamma(h_1\to W W^\star) =  \cos^2 \theta_{13}\ \Gamma(h^{\rm SM} \to W W^\star).
\label{eq:h1toww}
\eea
Using
\bea
 \Gamma(h^{\rm SM} \to \text{all}) &=& (4.07 +4\% -3.9\%) \,\text{GeV} \simeq 4.07 \pm 0.163\,\text{GeV}, \nn\\ 
  {\rm  and} \quad \Gamma(h^{\rm SM} \to \text{all})_{\rm LHC} &=& 3.7^{+1.9}_{-1.4}\,\text{GeV}
\eea
from reference \cite{ParticleDataGroup:2024cfk}, the corresponding  signal strength is given by
\bea
\mu_{{}_{WW^\star}}&=& \cos^4 \theta_{13}\ \frac{\Gamma(h^{\rm SM} \to \text{all})}{
	\Gamma(h^{\rm SM} \to \text{all})_{\rm LHC} } \simeq (1.1 \pm 0.57)   \cos^4 \theta_{13}
\label{eq:wwsignal}
\eea
Thus, the signal strength $\mu_{{}_{WW^\star}}$ depends only on the parameter $\theta_{13}$, which  can be strongly constrained by the observed value, from the LHC,   $\mu_{W W^\star} =  1.00 \pm 0.08$ \cite{ParticleDataGroup:2024cfk}.   Using the central value of  $\mu_{W W^\star} $ from \eqref{eq:wwsignal},  we get  \(0.92 \leq 1.1 \cos^4 \theta_{13} \leq1.08\), resulting in   the allowed range of $\theta_{13}$  to be 
\be 
5.5^\circ \leq \theta_{13} \leq 17^\circ.
\label{eq:theta13-range}
\ee 
%%%%%%%%%%%%%%%
\subsection{$h_1 \to\gamma \gamma$ and $h_1 \to Z \gamma$}
\label{subsec:muggZg}
The partial decay widths for $h_1 \to \gamma\gamma$ and $h_1 \to Z\gamma$ are generated at the one-loop level through the contributions of charged scalar $H^\pm$ and vector-like lepton $\chi$, and are parameterized in our model as
\begin{eqnarray}
\Gamma(h_1\to\gamma\,\gamma)&=& \cos^2\theta_{13} \left\vert 1 + \zeta_{\gamma\gamma}\right\vert^2\,\Gamma\left( h^{\rm SM} \to \gamma\,\gamma\right),
\label{h1toaa}\\
\Gamma(h_1\to Z\,\gamma)&=& \cos^2\theta_{13} \left\vert 1 + \zeta_{Z\gamma}\right\vert^2\,\Gamma\left( h^{\rm SM} \to Z\,\gamma\right).
\label{eq:h1toza}
\end{eqnarray}
where the SM Higgs partial decay widths in $\gamma\,\gamma$ and $Z\,\gamma$ channels are given as:
\begin{eqnarray}
	\Gamma(h^{\rm SM}\to\gamma\gamma)&=&\frac{G_F\alpha^2\ m_{h}^3}{128\sqrt{2}\pi^3}\ \left\vert  \frac{4}{3}  {\mathscr  M}^{\gamma\gamma}_{1/2}\bigg(\frac{4m_t^2}{m_{h^{\rm SM}}^2}\bigg)+  {\mathscr  M}^{\gamma\gamma}_1\bigg(\frac{4m_{\rm W}^2}{m_{h^{\rm SM}}^2}\bigg) \right\vert^2\label{hsmtoaa}\\
	\hspace{-5em} \Gamma (h^{\rm SM }\to Z\gamma ) &=& \frac{G^2_{F}\ \alpha\,m_W^2\, m_{h}^{3}} 
	{64\,\pi^{4}} \left( 1-\frac{m_Z^2}{m_{h^{\rm SM}}^2} \right)^3 \nn\\
	&&\times \left\vert
	2 \frac{\left( 1- \frac{8}{3} s^2_W \right)}{c_W} {\mathscr  M}^{Z\gamma}_{1/2}\left(\frac{4 m_t^2}{m_{h^{\rm SM}}^2},\frac{4 m_t^2}{m_Z^2}\right) + 
	{\mathscr  M}^{Z\gamma}_1\left(\frac{4 m_W^2}{m_{h^{\rm SM}}^2},\frac{4 m_W^2}{m_Z^2}\right) \right\vert^2. 
	\label{hsmtoza}
\end{eqnarray}
The dimensionless parameters $\zeta_{\gamma\gamma}$ and $\zeta_{Z\gamma}$ are defined as:
{\footnotesize
	\begin{eqnarray}
		\zeta_{\gamma\gamma} &=& \frac{v_{\rm SM}}{\cos\theta_{13}}\left[\frac{\frac{g_{h_1H^+H^-}}{2\,m^2_{H^\pm}} {\mathscr M}^{\gamma\gamma}_0\left(\frac{4 m_{H^\pm}^2}{ m_{h_1}^2} \right)  + \frac{y_2}{\sqrt{2} m_\chi} \sin\theta_{13} {\mathscr M}^{\gamma\gamma}_{1/2}\left(\frac{4\,m_{\chi}^2}{ m_{h_1}^2}\right)}{{\mathscr M}^{\gamma\gamma}_1\left(\frac{4\,m_{\rm W}^2} {m_{h_1}^2}\right) + \frac{4}{3}\, {\mathscr M}^{\gamma\gamma}_{1/2}\left(\frac{4\,m_t^2}{m_{h_1}^2}\right)}\right]\label{zetaaa}\\
		\zeta_{Z\gamma} &=& \frac{v_{\rm SM}}{\cos\theta_{13}}\left[\frac{-\frac{ g_{h_1 H^+ H^-}}{2 m_{H^\pm}^2} \frac{1- 2 s^2_{\theta_W}}{c_{\theta_W}} {\mathscr M}^{Z\gamma}_0\left(\frac{4 m_{H^\pm}^2}{m_{h_1}^2},\frac{4 m_{H^\pm}^2}{m_Z^2}\right) -  \frac{y_2}{\sqrt{2} m_\chi} \sin\theta_{13} \frac{4\ s^2_W}{c_W} {\mathscr M}^{Z\gamma}_{1/2}\left(\frac{4 m_\chi^2}{m_{h_1}^2},\frac{4 m_\chi^2}{m_Z^2}\right)}
		{2 \frac{\left( 1- \frac{8}{3} s^2_W \right)}{c_W} {\mathscr M}^{Z\gamma}_{1/2}\left(\frac{4 m_t^2}{m_{h_1}^2},\frac{4 m_t^2}{m_Z^2}\right) + {\mathscr M}^{Z\gamma}_{1}\left(\frac{4 m_W^2}{m_{h_1}^2},\frac{4 m_W^2}{m_Z^2}\right)}\right].
		\label{zetaza}
\end{eqnarray}}
where \(g_{h_1 H^+ H^-} = \vsm \lambda_{h_1 H^+ H^-}\).
 For the loop amplitudes ${\mathscr M}^{\gamma\gamma}_{0,\,1/2,\,1}$ and ${\mathscr M}^{Z\gamma}_{0,\,1/2,\,1}$, the reader is referred to the reference~\cite{Bharadwaj:2021tgp}. 
 Using the relations \eqref{eq:signalstrength}, \eqref{eq:h1toww},  and \eqref{eq:h1toza}, the ratios of signal strengths may be written as
 		\bea
 \frac{\mu_{{}_{\gamma \gamma(Z\gamma)}}}{\mu_{{}_{W \,W^\star}}}&=& \frac{\Gamma\left(h_1 \to \gamma \,\gamma(Z\gamma\right))}{\Gamma\left(h^{\rm SM} \to \gamma \,\gamma(Z\gamma)\right)} \times \frac{\Gamma(h^{\rm SM} \to W W^\star)}{\Gamma(h_1 \to W\, W^\star)} \, = \,   \left\vert 1 + \zeta_{\gamma\gamma(Z\gamma)} \right\vert^2
 \label{eq:signalratio}
 \eea
\section{Numerical Analysis}
\label{sec:num}
\subsection{Higgs Decay Signal Strengths}
The signal strengths \(\mu_{\gamma\gamma} \)  and  \(\mu_{Z\gamma} \) depend on five key model parameters: $\left\vert \theta_{13}\right\vert$,   $\lambda_{h_1H^+H^-}$, $y_2$, \mhpm, and $m_\chi$.  Among these,  $\theta_{13}$ is already constrained by  \(\mu_{WW^\star } \)  as specified  in equation~\eqref{eq:theta13-range}. The negative values of  $\theta_{13}$ are taken into account by changing the sign of the coupling  $y_2$ in the computation of both \(\mu_{\gamma\gamma} \)  and  \(\mu_{Z\gamma} \). Accordingly, we fix $\left\vert\theta_{13}\right\vert$ to three representative  values: \(5.5^\circ, \, 10^\circ\), and   \(15^\circ\),  and compute the one loop contributions to  \(\mu_{\gamma\gamma}\) and \(\mu_{Z\gamma}\)  by scanning the remaining parameters over the ranges: \( \lambda_{h_1H^+H^-} \in [-10:10], \,y_2 \in [-3.5:3.5] \), \(m_\chi  \in [200: 1000]\, \text{GeV}\), and \( \mhpm \in [\sqrt{m_{A^0}^2+m_{P^0}^2}: 1000]\, \text{GeV}\).  
\begin{figure}[h!]
			\centering
		\subfloat[{ \em{$y_2$ contours and density map in the $\mhpm-m_\chi$ plane }}\label{fig:muggzgy2} ]{
			\includegraphics[width=.6\textwidth]{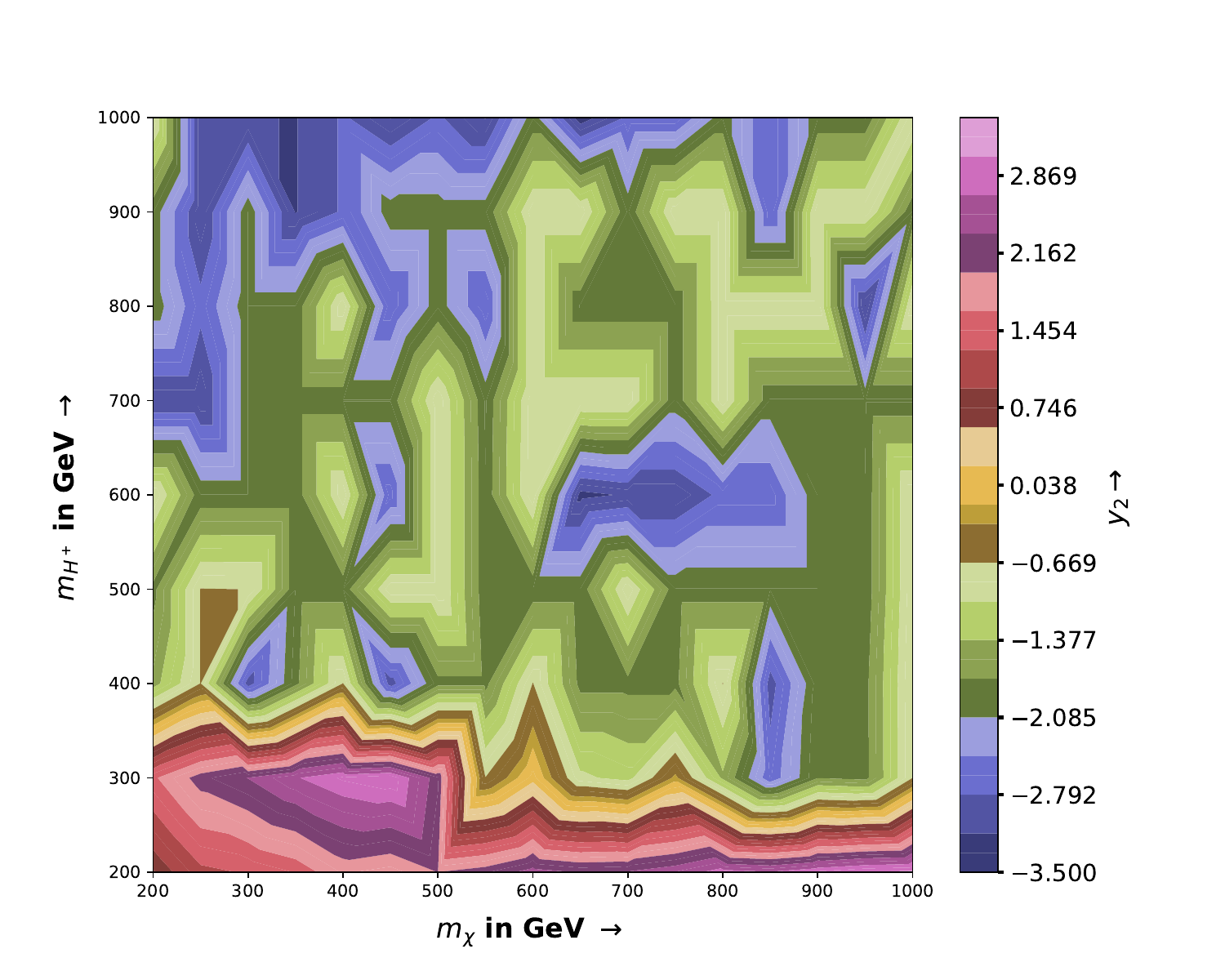}
		}\hfill
		\subfloat[{ \em{ $ \lambda_{{h_1 H^+ H^-}}$ contours and density map in the $\mhpm-m_\chi$ plane}}\label{fig:muggzglam} ]{%
			\includegraphics[width=.6\textwidth]{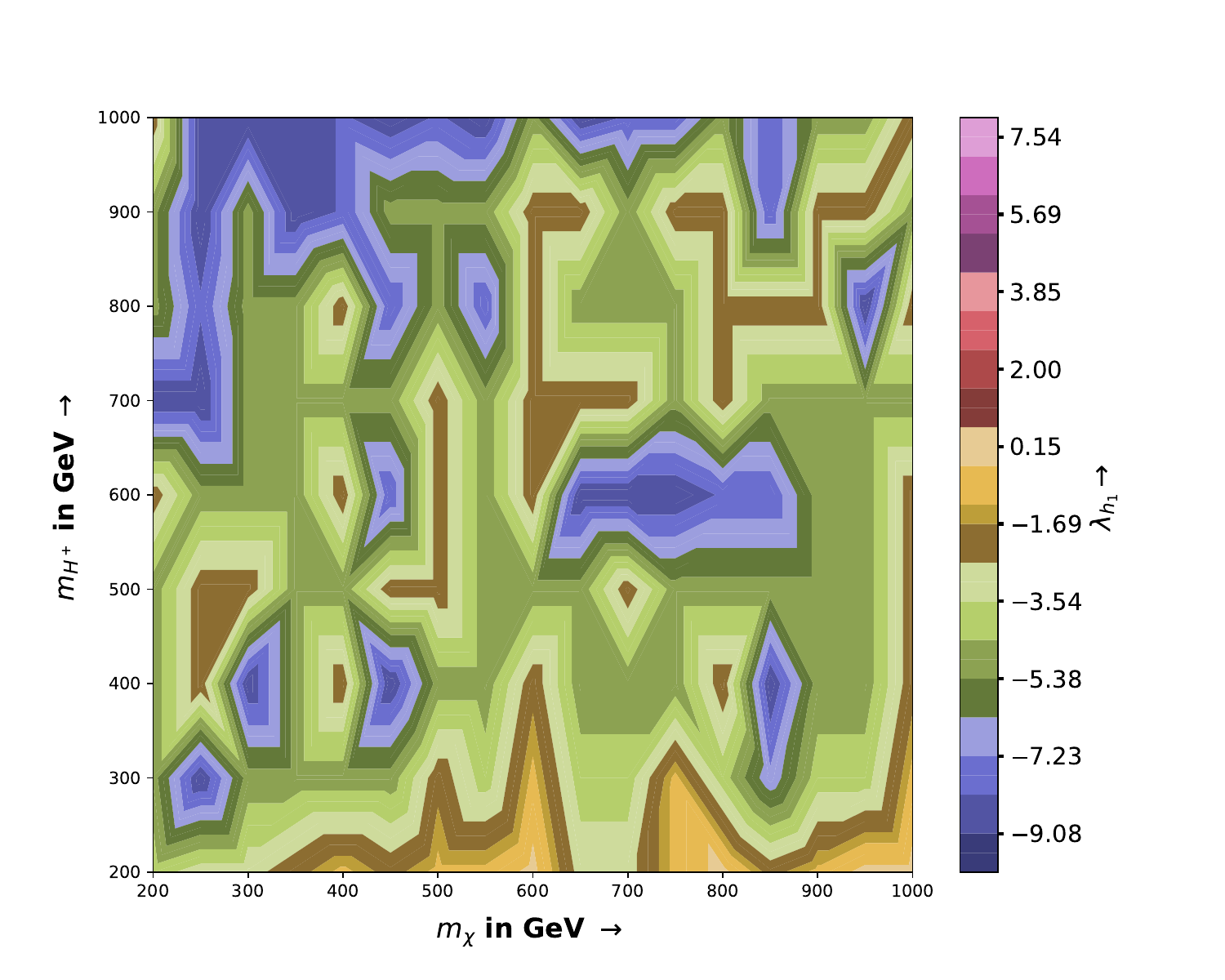}
		}\hfill
		\caption[nooneline]{\justifying \em{ Density maps of couplings $y_2$  and $\lambda_{h_1H^+H^-}$   in the $\mhpm-m_\chi$ plane. The contours represent parameter regions consistent with the observed Higgs decay signal strengths \(\mu_{Z\gamma}\) and \(\mu_{\gamma\gamma}\)  at the \(2\sigma\) level.}}
		\label{fig:hdecay}
	\end{figure}
Figure~\ref{fig:muggzgy2} (\ref{fig:muggzglam}) presents the density distribution of the  coupling $y_2$ ($\lambda_{{h_1 H^+ H^-}}$)  in the $\mhpm-m_\chi$ plane, consistent with the observed \(\mu_{\gamma\gamma}\) and \(\mu_{Z\gamma}\) signal strengths within the \(2\sigma\) experimental range. The color bar indicates the range of $y_2$ ($\lambda_{{h_1 H^+ H^-}}$), where positive and negative values are distributed across different mass regions. The structured density regions indicate strong correlations between these couplings and the Higgs decay constraints.
\begin{figure}[htb]
		\centering
			\includegraphics[width=.6\textwidth, height = 7cm ]{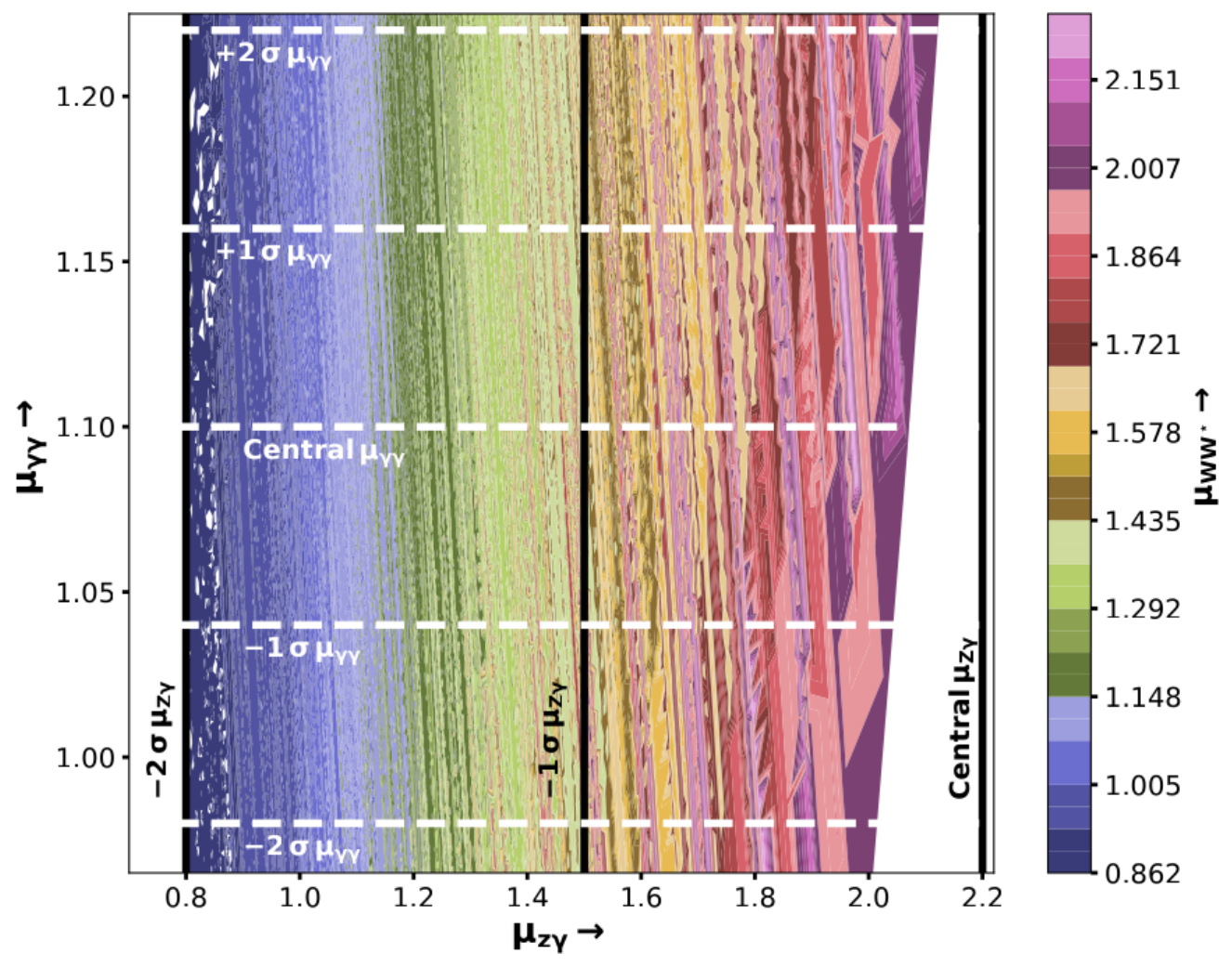}% 
		\caption[nooneline]{\justifying \em{Density plot of Higgs decay signal strength \(\mu_{WW}\) in  \(\mu_{Z\gamma}\)-\(\mu_{\gamma\gamma}\) plane. The white dashed lines denote the $\pm 1\sigma$ and $\pm2\sigma$ confidence levels for $\mu_{\gamma\gamma}$, with the central value highlighted. The pair of thick black  lines denote the  $- 2\sigma$ and $- 1\sigma$ values for \(\mu_{Z\gamma}\).}  }
\label{fig:muwwggzg}
\end{figure}
The figure~\ref{fig:muwwggzg} gives the parameter region consistent with all three signal strengths, namely,  \(\mu_{Z\gamma}\),  \(\mu_{\gamma\gamma}\)   and  \(\mu_{WW^\star}\),  at the  \(2\sigma\) level. We find that  \(\mu_{Z\gamma}\) is strongly dependent on 
\(\mu_{WW^\star }\),  while \(\mu_{\gamma \gamma}\) varies slowly with \(\mu_{WW^\star }\),  especially for lower values of \(\mu_{Z\gamma}\). Furthermore, a natural cut-off on \(\mu_{Z\gamma}\) appears for a given \(\mu_{\gamma \gamma}\) and \(\mu_{WW^\star }\),
as depicted by the right edge in figure~\ref{fig:muwwggzg}.  
%%%%%%%%%%%
\subsection{Model constraints}
\label{subsec:constraints}
We now proceed to constrain our model parameter space using (i) existing LEP II data, (ii) precision observables \(\Delta S\) and \(\Delta T\), and (iii) the muon anomalous magnetic moment.
%%%%%%%%%%
The  DELPHI and L3 combined analysis at LEP II ($\sqrt{s}=200$ GeV)  estimates the  muon pair production cross-section as~\cite{ALEPH:2013dgf}  
\be 
\sigma (e^+\ e^- \to \mu^+\ \mu^-)=3.072 \pm 0.108 \pm 0.018\, { \rm pb } .
\label{eq:mu-cross-LEP}
\ee
The excess contribution to this cross-section, beyond the SM prediction, arises from the $\left\vert y_1\right\vert$-induced Yukawa interactions between SM fermions and neutral scalars/ pseudoscalars \cite{Bharadwaj:2021tgp, Bharadwaj:2024gfo}. 

While LEP II constraints do not directly impact $\mu_{\gamma\gamma}$ and $\mu_{Z\gamma}$, they  significantly influence the parameters used to compute deviations $\Delta S$ and $\Delta T$, arising from radiative corrections due to extra scalars, pseudoscalars, and VLL contributions in SM gauge boson self-energies. The corrections are evaluated over a parameter space constrained by LEP II data and are required to lie within the \(1\sigma\)  experimental uncertainties:
\be 
\Delta S = -0.01 \pm 0.07,  \quad  \Delta T  = 0.04 \pm 0.06 \quad \cite{ParticleDataGroup:2024cfk}.
\label{eq:STU-values}
\ee
thereby imposing additional constraints on the model parameter space.
%%%%%%%%%%%%%
\begin{center}
	\begin{figure}[htb]
		\subfloat[{ \em{ $m_{A^0} = 150$ GeV, $m_{P^0} = 300$ GeV} }\label{fig:contour-r1}]{%
			\includegraphics[width=.49\linewidth]{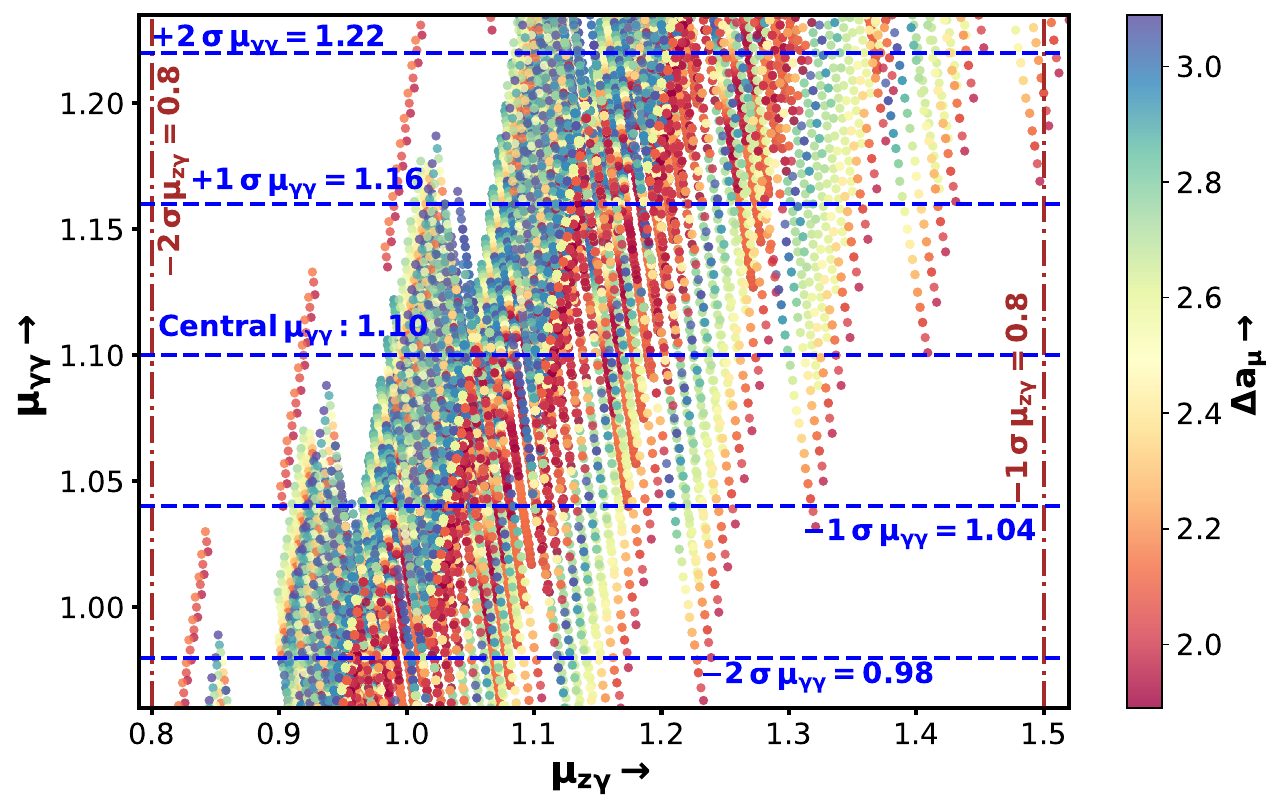}
		}\hfill%
		\subfloat[{ \em{ $m_{A^0} = 300$ GeV, $m_{P^0} = 150$ GeV}}\label{fig:contour-r-halfa}]{%
			\includegraphics[width=.49\linewidth]{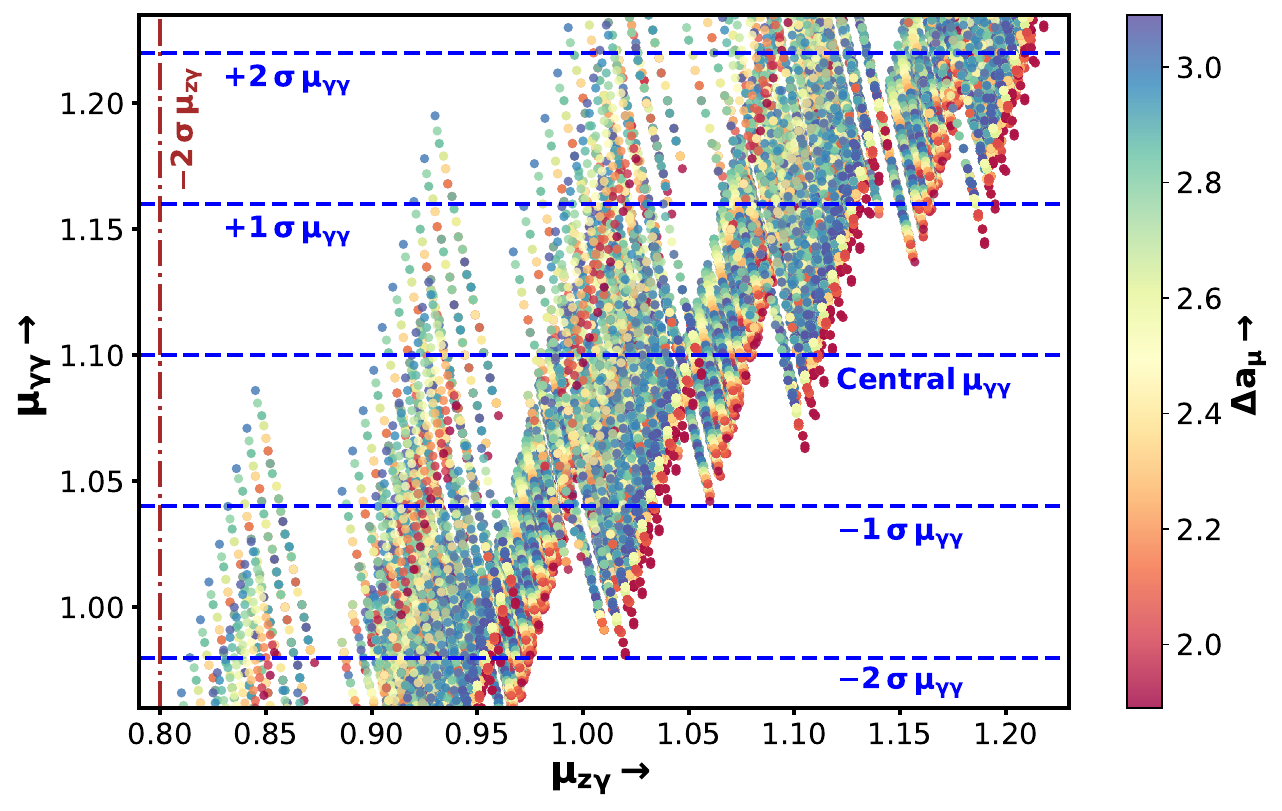}
		}\hfill%
		\subfloat[{ \em{ $m_{P^0} = m_{A^0} = 150$ GeV}}\label{fig:contour-r-halfb}]{%
			\includegraphics[width=.49\linewidth]{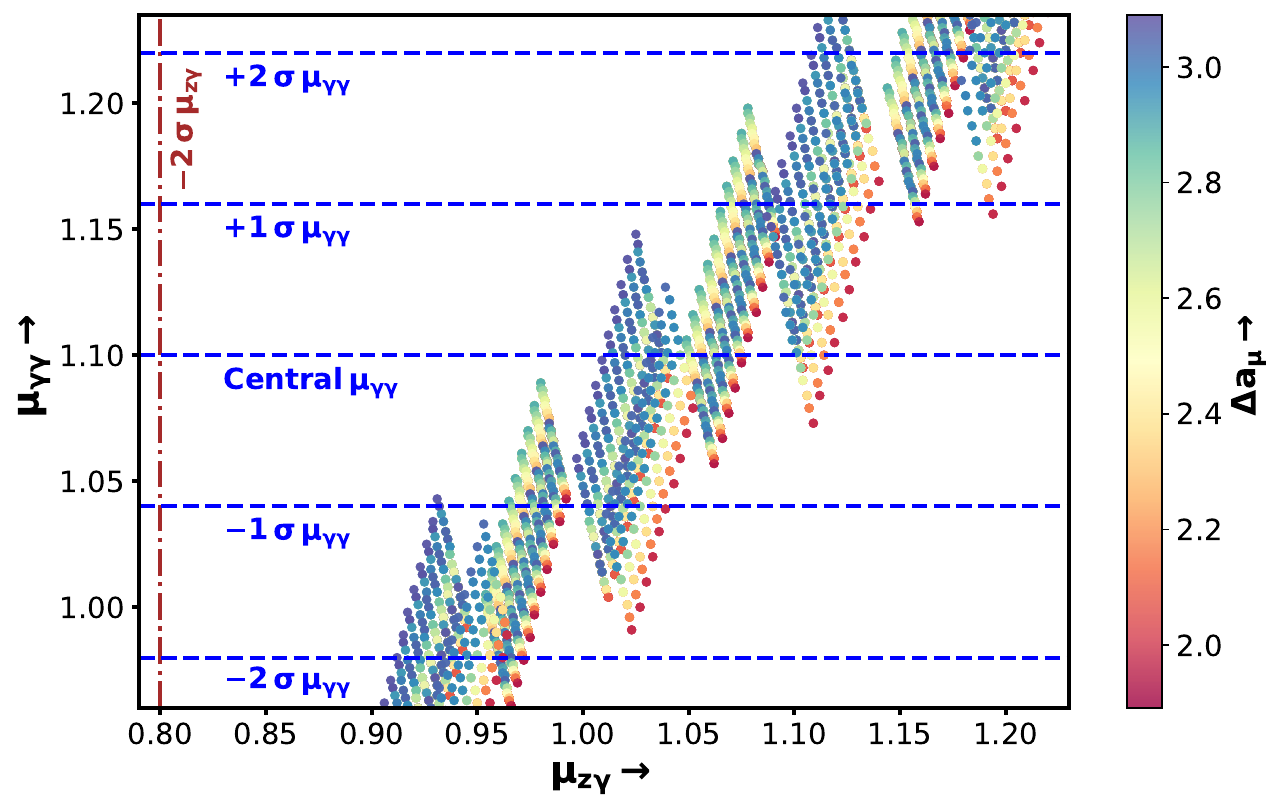}
		}\hfill%
		\subfloat[{ \em{$m_{A^0} = 150$ GeV, $m_{P^0} = 450$ GeV}}\label{fig:contour-r2}]{%
			\includegraphics[width=.49\linewidth]{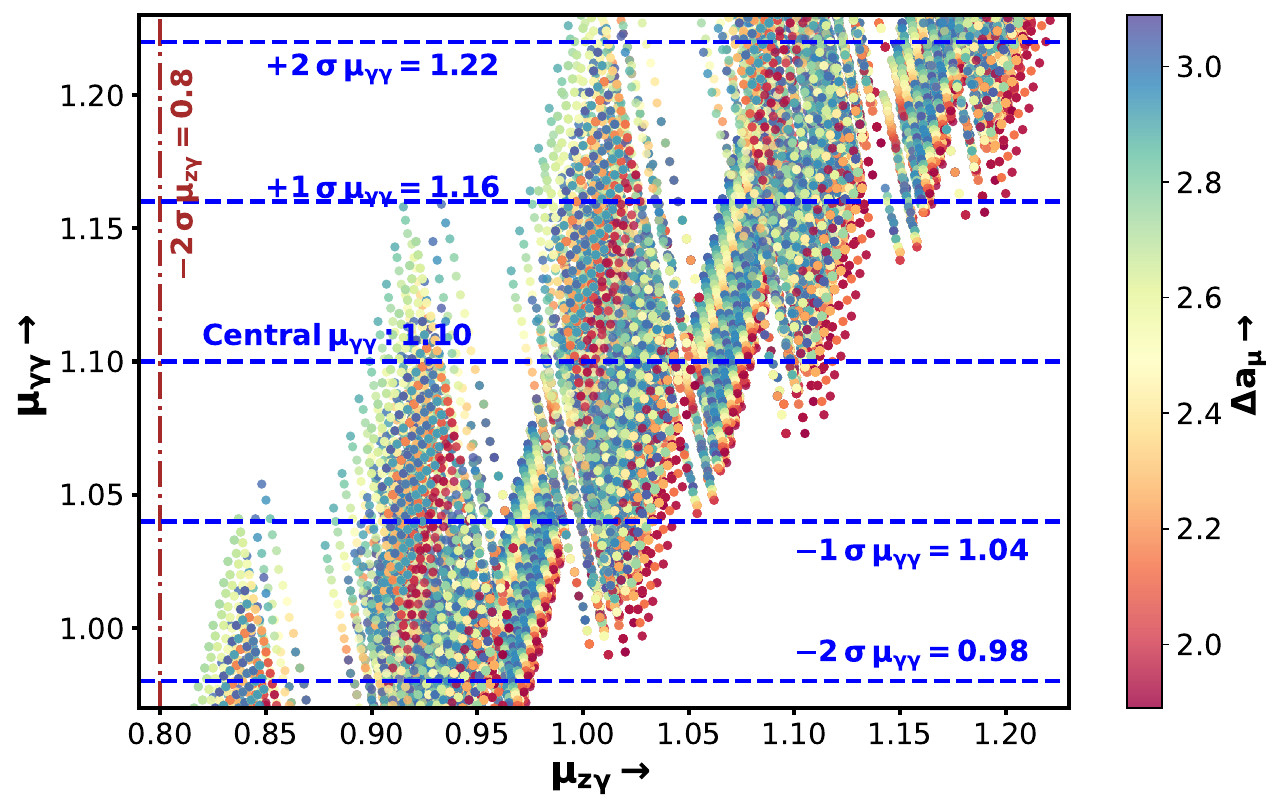}
		}\hfill%
		\caption[nooneline]{\justifying \em{ Scatter Plots depicting the combined solution consistent with signal strengths \(\mu_{Z\gamma}, \mu_{\gamma\gamma}\) at \(2\sigma\) level and muon \(g-2\) at \(1\sigma\) with the parameter space constrained by 	(i) \(\mu_{WW^*}\), (ii) \(\sigma (e^+\ e^- \to \mu^+\ \mu^-)\) from LEP, and  (iii) oblique parameters \(\Delta S,\, \Delta T\) at \(1\sigma\). The plots are shown for specific values of \(m_{A^0},\, m_{P^0}\) as mentioned. All other parameters are scanned in the range as given in section~\ref{subsec:constraints}.
		}}
		\label{fig:scatter}
	\end{figure}
\end{center}
%%%%%%%%
Next, we explore viable regions within the filtered model parameter sets to account for the observed anomalous magnetic dipole moment of the muon, as given by~\eqref{eq:MDMexpt}. To this end, we compute the one-loop and dominant two-loop Barr-Zee contributions to the muon’s anomalous magnetic moment within our model and subtract the SM contributions.

The resulting deviation, $\Delta a_\mu$,  arises from the exchange of additional VLLs as well as   charged and neutral scalars and pseudoscalars (for details, see reference ~\cite{Bharadwaj:2021tgp}). Requiring this excess contribution to remain within the $1\sigma$ uncertainty of the observed muon anomalous magnetic moment further shrinks  the model's parameter space. 

Finally, using the screened parameter sets, we compute \(\mu_{\gamma \gamma}\) and \(\mu_{Z \gamma}\) demonstrating the simultaneous explanation of both signal strengths while satisfying all the low energy constraints.

Due to paucity of space, we illustrate the existence of a viable region by selecting four sets of $\left(m_{A^0},\,m_{P^0}\right) $: $\left(150,\, 300\right)$ GeV,  $\left(300,\, 150\right)$ GeV, $\left(150,\, 150\right)$ GeV, and $\left(150,\, 450\right)$ GeV. For each of these choices, we scan the remaining parameters to compute the $\mu_{\gamma\gamma}$ and $\mu_{Z\gamma}$, ensuring consistency with key low-energy constraints as discussed in the previous section. The parameters are varied within the following ranges:
\begin{itemize}
\item Mixing Angles:  \hskip 0.5cm  $\left\vert\theta_{13}\right\vert=5.5^\circ, 10^\circ$,  and  $15^\circ $, alongwith 
\(15^\circ  \leq \left\vert\theta_{23}\right\vert \leq 90^\circ\).\\
Note that negative values of  $\theta_{13}$ and $\theta_{23}$ are effectively incorporated by appropriately adjusting the signs of the couplings $\lambda_{h_1H^+H^-}$ and $y_2$ in the computation of both the Higgs signal strengths and the anomalous magnetic moment of muon.
\item Couplings:   \hskip 0.5cm 
\(-10 < \lambda_{{h_1 H^+ H^-}} < 10\),  \,\,\,\,\,\,\, {\rm and}  \,\,\,\,\,\,\, \(-10 < \lambda_{{h_3 H^+ H^-}} < 10\).  
\item Masses:\\  (i) \( \sqrt{m_{A^0}^2 + m_{P^0}^2} < m_{h_2},\, \mhpm < 1000\) GeV,\,\,\,\, (ii) \( 250 <  m_{h_3} < 1000\) GeV, \,\,\, and \,\,\, (iii) \( m_\chi \in [200, 1000]\) GeV.
\end{itemize} 
The results of this analysis are encapsulated in Figure~\ref{fig:scatter}, where the
four scatter plots display the viable regions in \(\mu_{Z\gamma}\) - \(\mu_{\gamma \gamma}\)  plane.  These regions satisfy the $2\sigma$ bounds from the combined CMS and ATLAS analyses while also incorporating constraints from the Higgs signal strength $\mu_{WW^\star}$, LEP bounds on $\sigma\left(e^+e^-\to\mu^+\mu^-\right)$, and the electroweak oblique parameters \(\Delta S,\,\, \Delta T\). Each panel corresponds to a distinct benchmark choice of $\left(m_{A^0},\,m_{P^0}\right)$,  while the color bar encodes the contribution to the muon anomalous magnetic moment  $\Delta a_\mu$, constrained within its $1\sigma$ experimental range. We observe   that 
\begin{itemize} 
\item The viable parameter space forms structured, diagonal bands, indicative of a strong correlation between \(\mu_{Z\gamma}\) and \(\mu_{\gamma \gamma}\), which  stem from shared underlying parameter dependencies. 
\item The high-density regions in the scatter plots, where numerous viable points cluster, indicate parameter configurations that most consistently satisfy all experimental constraints, highlighting the most robust areas of the model’s viable parameter space.
\item The solutions tend to cluster near the central value of $\mu_{\gamma\gamma}$, while   showing a preference for $\mu_{Z\gamma}\lesssim 1.5$, which lies below the current central experimental value of 2.2. Should future measurements shift the experimental $\mu_{Z\gamma}$ closer to the SM prediction, the model's viability would be further strengthened.
 \end{itemize}
\section{Summary }
\label{sec:sum}
We have investigated the parameter space of a constrained 2HDM  extended by a complex scalar singlet and a VLL, incorporating constraints from LEP II data, electroweak precision observables ($\Delta S$, $\Delta T$), and the muon anomalous magnetic moment $\Delta a_\mu$.

A detailed numerical scan over the model parameters, including mixing angles, scalar quartic couplings, and mass eigenvalues within experimentally allowed intervals, reveals sizable regions of the model parameter space that are consistent with the observed Higgs boson decay signal strengths $\mu_{\gamma\gamma}$ and $\mu_{Z\gamma}$ within their two sigma experimental ranges. As shown in Figure~\ref{fig:scatter}, these viable regions also satisfy the constraints from the muon $(g-2)_\mu$ anomaly, along with electroweak precision bounds and LEP II exclusion limits. These results demonstrate that the model offers a phenomenologically viable framework capable of addressing current experimental anomalies while maintaining consistency with all relevant precision observables.

\section*{Acknowledgements}
We acknowledge partial financial support from the ANRF grant CRG/2023/008234. S.D. further acknowledges support from the project grant IOE/2024-25/12/FRP of the University of Delhi.

\end{document}